\documentclass[aps,prl,twocolumn,showpacs,superscriptaddress]{revtex4-1}

\usepackage[]{graphicx}
\usepackage{times, units}
\usepackage{bm}
\usepackage{blindtext}
\usepackage{color}

\usepackage{amsmath}
\usepackage{amssymb}
\usepackage{array}

\newcommand{\comment}[1]{}

\newcommand{\fett}[1]{{\boldsymbol{ #1}}}
\newcommand\mat[1]{\mathcal{#1}}

\begin{document}

\title{Impact of strain on the optical fingerprint of monolayer transition metal dichalcogenides}
\author{Maja Feierabend}
\affiliation{Chalmers University of Technology, Department of  Physics, 412 96 Gothenburg, Sweden}
\author{Alexandre Morlet}
\affiliation{\'{E}cole Normale Supérieure de Cachan, D\'{e}partement de Physique, 94230 Cachan, France}
\author{Gunnar Bergh\"auser}
\affiliation{Chalmers University of Technology, Department of  Physics, 412 96 Gothenburg, Sweden}
\author{Ermin Malic}
\affiliation{Chalmers University of Technology, Department of  Physics, 412 96 Gothenburg, Sweden}

\begin{abstract}
Strain presents a straightforward tool to tune electronic properties of atomically thin nanomaterials that are highly sensitive to lattice deformations. While the influence of strain on the electronic band structure has been intensively studied, there are only few works on its impact on optical properties of monolayer transition metal dichalcogenides (TMDs). Combining microscopic theory based on Wannier and Bloch equations with nearest-neighbor tight-binding approximation, we present an analytical view on how uni- and biaxial strain influences the optical fingerprint of TMDs including their excitonic binding energy, oscillator strength, optical selection rules, and the radiative broadening of excitonic resonances. We show that the impact of  strain can be reduced to changes in the lattice structure (geometric effect) and in the orbital functions (overlap effect). 
 In particular, we demonstrate that the valley-selective optical selection rule is softened in the case of uniaxial strain due to the introduced asymmetry in the lattice structure. Furthermore, we reveal a considerable increase of the radiative dephasing due to strain-induced changes in the optical matrix element and the excitonic wave functions.
\end{abstract}

\maketitle
Atomically thin transition metal dichalcogenides (TMDs) have been in the focus of current research due to their 
efficient light-matter interaction and the remarkably strong Coulomb interaction  leading to tightly bound excitons  \cite{He2014,gunnar_prb,Chernikov2014,Ramasub2012}. Recently, the impact of strain on optical and electronic properties of TMDs has gained importance, since these atomically thin materials are highly sensitive to deformations of their lattice structure suggesting strain-induced tailoring of TMD characteristics. Recent experimental \cite{he2013experimental,conley2013bandgap,feng2012strain,island2016precise,schmidt2016reversible} 
and theoretical \cite{steinhoff2014influence,scalise2014first,san2016inverse,horzum2013phonon,yue2012mechanical, shi2013quasiparticle,amin2014strain,wang2014many,schmidt2016reversible} studies have revealed that strain can significantly change the electronic band structure of TMDs. In particular, the direct band gap decreases (increases) for tensile (compressive) strain resulting in a considerable red (blue) shift of optical resonance. 
So far, most theoretical studies on the impact of strain in TMDs are based on DFT calculations focusing on changes in the electronic band structure without taking into account the predominant role of excitons in these materials. 
 In this work, we present an analytic approach combining the Wannier and TMD Bloch equations for excitons with the nearest-neighbor tight-binding wave functions. The goal is to provide a microscopic access to the impact of uni- and biaxial strain on the optical fingerprint of TMDs including the excitonic binding energy, the oscillator strength, the optical selection rules, and the radiative broadening of excitonic resonances.

\begin{figure}[t]
  \begin{center}
\includegraphics[width=\linewidth]{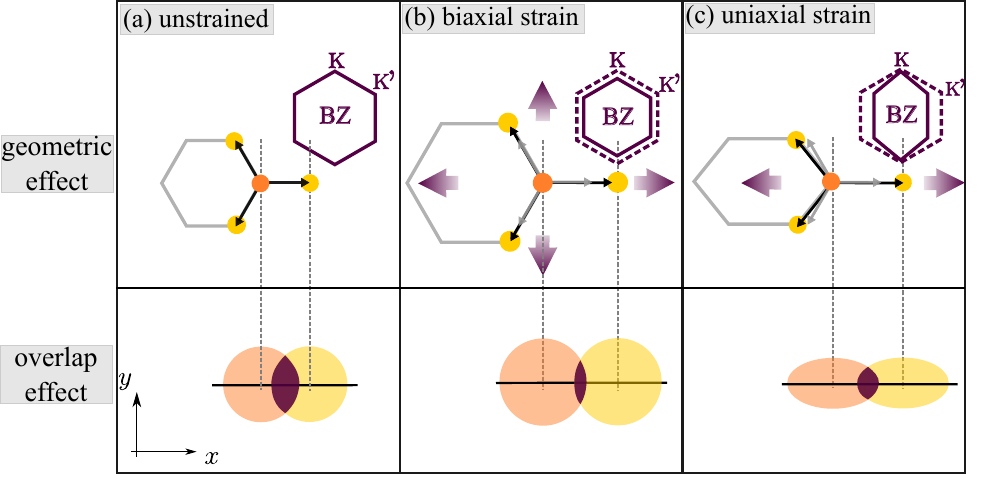} 
\end{center}
    \caption{ 
Strain affects electronic and optical properties of transition metal dichalcogenides MX$_2$ (with a transition metal M and two chalcogen atoms X) through (i) geometric changes in the real space lattice (geometric effect) and (ii) changes in the orbital functions (overlap effect), upper and lower panel respectively.
(a) The upper panel shows the hexagonal lattice structure with M (orange) and X (yellow) atoms in real space and the corresponding Brillouin zone (BZ) in momentum space in the unstrained case. The lower panel represents the corresponding orbitals functions  (orange for the M orbitals, yellow for X orbitals) and their overlap (purple). 
(b) In presence of tensile biaxial strain, atoms are uniformly moved apart in both directions. Hence, the hexagonal lattice structure remains symmetric. In momentum space, this leads to a decrease of the BZ size. Furthermore, due to the larger distance between M and X atoms, the orbital function overlap is reduced.
(c) In the case of tensile uniaxial strain, i.e. strain only along one direction (here x), the hexagonal structure becomes antisymmetric both in real and momentum space. Beside the reduced orbital overlap, the atomic orbital functions also become elliptic in the direction of strain. 
}
  \label{schema}
\end{figure}

In a first step, we determine the electronic band structure of a general TMD material MX$_2$ with M=(Mo, W) and X=(S, Se), where Mo and W stand for molybdenum and  tungsten transition metals, while S and Se denote the sulfur and selenium chalcogen atoms. To this end, we solve the stationary Schr\"{o}dinger equation $H\Psi^{\lambda}_{{\bf k}}({\bf r}) = E^{\lambda}_{{\bf k}} \Psi^{\lambda}_{\bf k}(\bf r)$
including the tight-binding (TB) wave functions 
$
\Psi^{\lambda}_{\bf k} ({\bf r}) = \frac{1}{\sqrt{N}} \sum_{j=\text{M, X}} C^{\lambda}_{j\bf k} \sum_{R_j}^N  e^{i \bf k \cdot R_j} \phi^{\lambda}_j ({\bf r - R_j }).
$
Here, the TB coefficients $C^{\lambda}_{j\bf k}$ express the contribution of each atomic sublattice $j=$(M,X),  $\bf R_j$ represents the position of the atoms within the sublattice $j$ and $\phi^{\lambda}_j ({\bf r - R_j })$ stands for the atomic orbitals that are relevant for the considered bands $\lambda$.  By inserting the nearest-neighbor TB approximation \cite{carbonbuch, reich2002tight},  we obtain an expression for the electronic band structure
\begin{equation}
\label{eq_E0}
E^{\lambda}_{\bf k}=\pm
\frac{1}{2} \sqrt{
E_{\text{gap}}^2 + 4|t^{\lambda}|^2 \,|e({\bf k}) | ^2
}.
\end{equation}
The electronic band gap $E_{\text{gap}} = \frac{1}{2}\sum_{\lambda}\left(H_{ii}^{\lambda} - H_{jj}^{\lambda}\right)$ is determined by the on-site energies $H_{ii}^\lambda$. Furthermore, the nearest-neighbor hopping integral reads $H_{ij}^{\lambda} =\langle \phi^{\lambda}_i |H |\phi^{\lambda}_j \rangle= t^{\lambda} e({\bf k})=t^{\lambda} \sum_\alpha e^{-i\bf k b_\alpha} $
with  $
t^{\lambda} =
\langle \phi^{\lambda}_i ({\bf r - R_i }) |H| \phi^{\lambda}_j ({\bf r - R_j }) \rangle 
$  and the nearest-neighbor connecting vectors $\bf b_\alpha$.
To obtain Eq. (\ref{eq_E0}), we have exploited the symmetry  of the lattice resulting in $H_{ij}^{\lambda} = H_{ji}^{\lambda *} $ and neglecting the overlap of orbital functions of neighboring sites, i.e. $S_{ij} = \langle \phi^{\lambda}_i |\phi^{\lambda}_j \rangle= \delta_{ij}$.
Finally, restricting our investigations to the area around the high-symmetry K point in the Brillouin zone, we can further simplify the electronic band structure by performing a Taylor expansion for small momenta $\bf k$ and by applying the effective-mass approximation:
\begin{equation}
\label{eq_E}
E^{\lambda}_{\bf k} \approx \sigma_\lambda \left(
\frac{E_{{gap}}}{2} + \frac{\hbar^{2}}{2m^{\lambda}} {\bf k}^2 
\right)
\end{equation}
with $\sigma_c=+1, \sigma_v=-1$ and with the effective mass  
\begin{equation}
\label{eq_m}
m^{\lambda}= \frac{2 \hbar^2E_{\text{gap}}}{3\,|t^{\lambda}|^2},
\end{equation}
that is given by the TB hopping parameter $t^{\lambda}$. The latter determines the curvature of the electronic bands around the K point. Solving the Schr\"{o}dinger equation, we also obtain the TB coefficients:
\begin{equation}
\label{eq_C}
  C^{\lambda}_{X\bf k} = \left(1+|g^{\lambda}_{\bf k}|\right)^{-\frac{1}{2}}, \quad C^{\lambda}_{M\bf k} =  C^{\lambda}_{X\bf k} \,g^{\lambda}_{\bf k}  
\end{equation}
with $g^{\lambda}_{\bf k} = t^{\lambda} \, e({\bf k}) \left(\frac{E_{\text{gap}}}{2} - E^{\lambda}_{\bf k}\right)^{-1}$.

Now, we have access to the electronic band structure and the electronic eigen function of unstrained TMDs. Putting these materials under strain leads to two effects having impact on electronic and optical properties of TMDs: (i) geometric effect  and (ii) orbital overlap effect, cf. Fig. \ref{schema}. 
The geometric effect describes the change in the geometry of the lattice compared to the unstrained case (Fig. \ref{schema}(a)). For biaxial strain, i.e. strain applied both to x and y direction,  this simply implies an increase in the lattice constant $a_0$ (Fig. \ref{schema}(b)), while for uniaxial strain, i.e. strain applied only in one direction, the lattice vectors change differently in both directions leading to a broken lattice symmetry  (Fig. \ref{schema}(c)). 
In momentum space, the Brillouin zone changes accordingly: biaxial strain leads to an uniform decrease of the zone, while uniaxial strain implies a compression only in the direction of the applied strain. 
Beside the pure geometric effect, strain has also an effect on the overlap of the atomic orbitals, cf. the lower panel of Fig. \ref{schema}. Here, the crucial property is the overlap of M and X  orbital functions. Due to the  strain-induced increase  in the distance between atoms, the overlap of the orbitals is reduced. The effect is more pronounced for biaxial strain, while for uniaxial strain the broken lattice symmetry is transferred to the orbital shape (elliptic form). In this work, we focus on tensile strain, however the gained insights can be also applied to compressive strain.

We implement the geometric effect of the strain  by introducing the strain matrix 
$\mat{S}= \left(\begin{array}{cc}
s_x & 0 \\
0 & s_y \end{array} \right)$,
where $s_{i}=1 \pm \frac{s [\%]}{100 \%}$
with $s$ denoting the strain value and $\pm$ representing tensile and compressive strain, respectively. 
The basis vectors in real (momentum) space transform to  $\fett{a}_i \rightarrow \mat{S}\fett{a}_i$ ($\fett{k}_i \rightarrow \mat{S}^{-1} \fett{k}_i$) which leads to an increase (decrease) of the hexagonal lattice, cf. the upper panel in Fig. \ref{schema}.  We neglect changes in the z-direction, since the effect has been recently shown to be negligibly small \cite{ wang2014many}.\\
Within the nearest-neighbor TB approximation  the atomic orbitals $\phi_{j}^{\lambda}$ appear in integrals of the form $\langle \phi_{i}^{\lambda}|H|\phi_{j}^{\lambda} \rangle$. 
Since we are not interested in the exact shape of the orbitals, but only in their strain-induced change, we assume effective 1s hydrogen-like atomic orbitals
$
\phi_{j}^{\lambda}({\bf r})=N_{j} \exp{(-\frac{{\bf r}-\mat{S}{\mathbf R}_j}{\sigma_{j, \lambda}})}
$
with the normalization constant $N_j$, the atomic positions ${\mathbf R}_j$, and the orbital width $\sigma_{j, \lambda}$. We allow the width to change with strain and find a self-consistent solution by benchmarking the theory to experimentally observed strain-induced shifts in optical spectra \cite{he2013experimental,conley2013bandgap}. 
Inserting this ansatz in Eq. \eqref{eq_E} we can find analytic expressions for the strain-dependent electronic band gap $E_{\text{gap}}(s)$ and the TB hopping integral $t^{\lambda}(s)$ in the case of symmetric biaxial strain ($s_x=s_y=s$):
\begin{eqnarray}
\label{eq_Egap}
 E_{\text{gap}}(s) &=&    \frac{\hbar^2}{4\,m_0 \,s^{2}} 
\sum_{\lambda} \left(
\sigma_{i,\lambda}^{-2} - 
\sigma_{j,\lambda}^{-2}
\right), \\
t^{\lambda}(s) &=&  \frac{2\hbar\, s}{m_0 (\sigma_{i,\lambda} + \sigma_{j,\lambda})^2}. \label {eq_t}
\end{eqnarray}
Both quantities show a clear dependence on strain predominantly via the orbital overlap effect (reflected by $\sigma_{i,\lambda},\sigma_{j,\lambda}$): the band gap decreases with $s^{2}$, while the hopping integral linearly increases with $s$. The slope of the increase/decrease depends on the widths of the atomic orbitals and is hence TMD specific -  in agreement to experimental observations, where we find a band gap reduction of approximately 50 meV/$\%$ strain in WSe$_2$ and MoS$_2$ \cite{schmidt2016reversible, he2013experimental}.  We exploit these experimental findings to benchmark our theory by adjusting the widths and the overlap of the atomic orbitals. 

\begin{figure}[t]
  \begin{center}
\includegraphics[width=\linewidth]{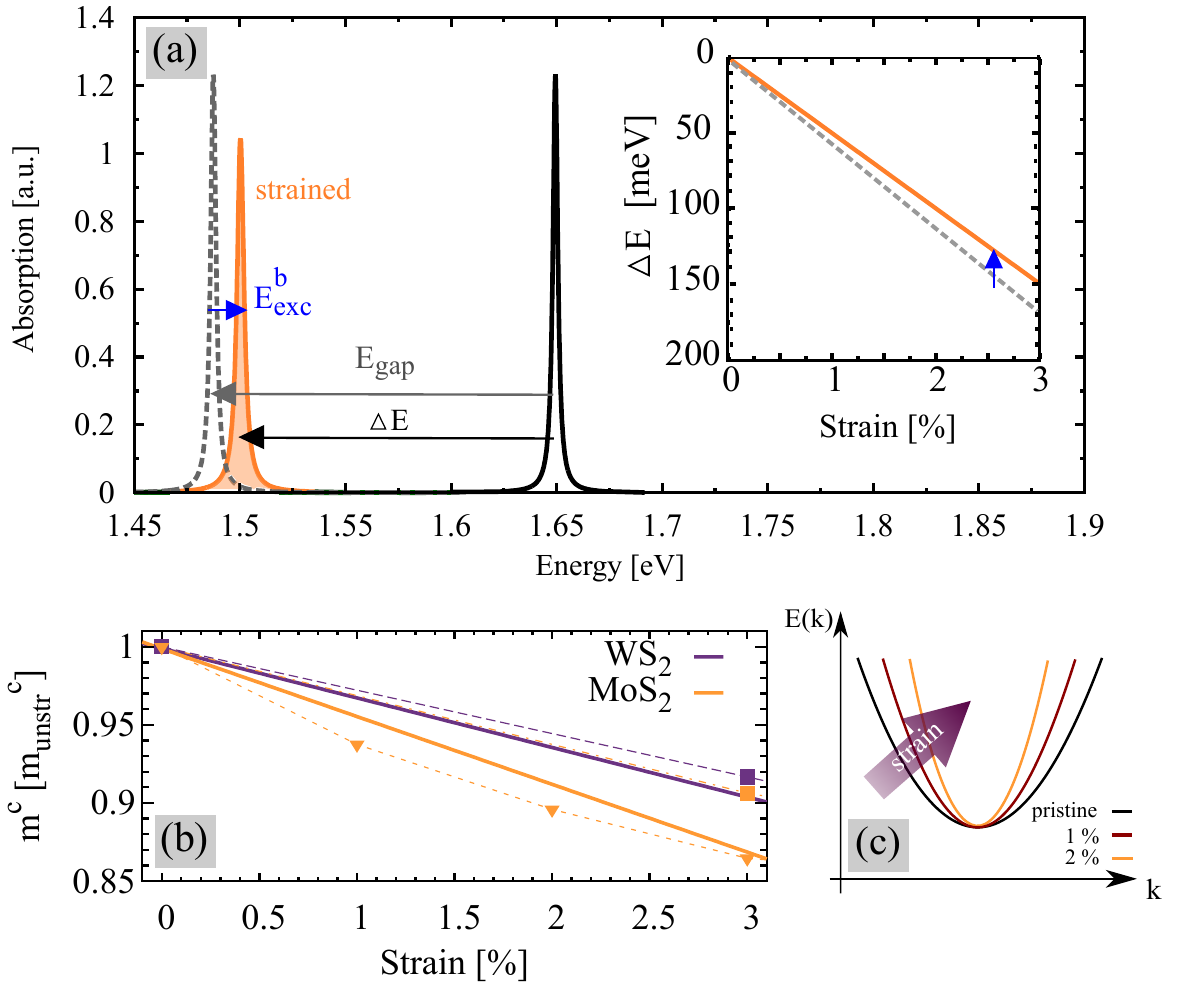} 
\end{center}
    \caption{  (a) Excitonic absorption spectra of unstrained (black) and uniaxially strained tungsten diselenide (WSe$_2$) as exemplary TMD material at 3 $\%$ strain. The observed red-shift stems from  (i) a decrease in the orbital overlap giving rise to a reduced band gap ($E_{\text{gap}}$) and hence a red-shift (dashed gray line) and (ii) the geometric effect leading to a decrease in the effective masses, which results in weaker bound excitons ($E^b_{exc}$) and hence a blue-shift of the unstrained peak.  The inset shows the resulting energy shift $\Delta E$ as a function of strain both with (orange) and without (gray dashed) taking into account excitonic effects.
(b) Strain-dependent decrease of the effective mass in the conduction band for WS$_2$ (purple) and MoS$_2$ (orange). Our results (solid lines) are in good agreement with values obtained by DFT calculations (symbols and dashed lines)  taken from Ref. \cite{yue2012mechanical} (triangles) and Ref. \cite{shi2013quasiparticle} (squares). (c) Sketch of the effect of strain on the dispersion of the conduction band.
}
  \label{excAbsorption}
\end{figure}

The strain-induced change of the band gap (Eq. \eqref{eq_Egap}) gives rise to a red-shift of electronic resonances in optical spectra, cf. the dashed gray line in Fig. \ref{excAbsorption} (a). The strain-induced change of the TB hopping integral (Eq. \eqref{eq_t}) determines the variation of the effective mass, i.e. the inverse band curvature. We observe a clear decrease in the effective mass of the conduction band (Fig. \ref{excAbsorption} (b)) and a corresponding increase in the band curvature (Fig. \ref{excAbsorption} (c)). We predict a reduction of the effective mass by 3$\%$ for WS$_2$ (purple line) and 4$\%$ for MoS$_2$ (orange line) in the case of 1$\%$ applied uniaxial strain. In case of biaxial strain, the effect is roughly twice as large. For the effective mass in the valence band, we obtain similar results.
Our results are in good agreement with DFT calculations (dashed lines) \cite{yue2012mechanical, shi2013quasiparticle}.

By performing these studies, we have benchmarked our theory with available experimental and DFT studies regarding the impact of strain on the electronic properties of TMDs (band gap and band curvature). Now, we include excitonic effects and investigate how they change in presence of bi- and uniaxial strain. Excitons are integrated by solving the  Wannier equation providing access to eigenvalues and eigen functions for all available excitonic states \cite{Kochbuch,Kira2006,kuhn2004, gunnar_prb, feierabend17}. 
Furthermore, we derive the TMD Bloch equation for the microscopic polarization $
 p_{\bf{k_1, k_2}}^{vc}(t) =\langle a^+_{c,\bf{k_1}}a^{\phantom{+}}_{v,\bf{k_2}} \rangle (t)
$ giving access to excitonic optical response of TMDs \cite{carbonbuch}. 
This microscopic quantity is a measure for optically induced transitions from the state $(v,\bf{k})$ to $(c,\bf{k})$ that are characterized by the electronic momentum $\bf{k_i}$ and the band index $\lambda_i=(v,c)$ denoting the valence and the conduction band, respectively \cite{carbonbuch}. 

Since excitonic effects are known to dominate optical properties of TMDs \cite{gunnar_prb,berkelbach,RIS_0}, we project the microscopic polarization into an excitonic basis \cite{thranhardt2000quantum}
$
 p_{\bf{k_{1},k_{2}}}^{vc}(t) \rightarrow p_{\bf{qQ}}^{v c} (t)= \sum_{\mu} \varphi_{\bf q }^{\mu} p_{\bf Q}^{\mu} (t)
$
with excitonic eigen functions $\varphi_{\bf q }^{}$ and the index $\mu$ representing the excitonic state. In this work, we focus on the energetically lowest optically allowed $A_{1s}$ state. Furthermore, we introduce center-of-mass  and  relative momenta $\bf{Q}$ and $\bf{q}$, where $\bf{Q=k_{2}-k_{1}} $ and ${\bf q}=\frac{m_{h}}{M}{\bf k_{1}}+\frac{m_{e}}{M} {\bf k_{2}}$ with the electron (hole) mass $m_{e(h)}$ and the total mass $M=m_e+m_h$.
 The separation ansatz enables us to decouple the relative from the center-of-mass motion. For the relative coordinate including the reduced mass $\mu = \frac{m_cm_v}{m_c+m_v}$, we solve the Wannier equation \cite{Kochbuch,Kira2006,kuhn2004, gunnar_prb}
 \begin{equation}\label{eq_wannier}
 E_{{\bf q}}  \varphi_{{\bf{q}}}
 - \sum_{{\bf{k}}} V_{\text{exc}}({\bf{k}})  \varphi_{{\bf{q-k}}} = E_{\text{exc}}^{\text{b}} \varphi_{\bf{q}}
\end{equation}
with  the excitonic eigen function $\varphi_{\bf {q}}^{}$, the excitonic binding energy  $E_{\text{exc}}^{\text{b}}$, and the dispersion $E_{{\bf q}}=\frac{\hbar^2 q^{2}}{2\mu}$.

To obtain the temporal evolution of the excitonic microscopic polarization $p_{\bf Q}(t)$, we solve the Heisenberg equation of motion $i\hbar \dot p_{\bf Q} (t)=[H,p_{\bf Q} (t)]$ \cite{Kochbuch, carbonbuch}. This requires the knowledge of the many-particle Hamilton operator $H = H_0 + H_{c-l} + H_{c-c}$ including the free carrier contribution $H_0$, the carrier-light interaction $H_{c-l}$ and the carrier-carrier interaction $H_{c-c}$. 
To calculate the coupling elements, we apply the nearest-neighbor tight-binding approach \cite{carbonbuch,Kochbuch,Kira2006}.
Exploiting the fundamental commutator relations \cite{Kochbuch}, we obtain the TMD Bloch equations for the excitonic microscopic polarization \cite{gunnar_prb}:
\begin{eqnarray}
\label{eq_bloch}\dot p_{\bf{Q}}(t)
&=&
\frac{1}{i\hbar}\left(E_{\text{exc}} + 
\frac{\hbar^2 Q^{2}}{2M}
-i\gamma\right)\,
p_{\bf{Q}}(t)
 +
\Omega(t)\, \delta_{\bf{Q,0}} 
 \end{eqnarray}
The optical excitation is expressed by the Rabi frequency 
\mbox{
$
\Omega(t)=\frac{e_{0}}{m_0}  
\sum_{\bf{q}} \varphi_{\bf q}^{ *} 
\boldsymbol M^{vc}({\bf {q}})\cdot \boldsymbol A(t)
$}
including the optical matrix element $\boldsymbol M^{vc}({\bf {q}})$ and the external vector potential $\boldsymbol A(t)$. Here,  $e_0$ denotes the electron charge and $m_0$ the electron rest mass, respectively. 
Taking into account only direct transitions with $\bf{Q}=0$, we end up in an analytic expression for the absorption coefficient corresponding to the well-known Elliot formula \cite{Kira2006,Kochbuch}: 
\begin{equation} \label{eq_elliot}
\alpha(\omega) \propto \frac{1}{\omega} \Im \left[ \frac{|\sum_{{\bf{q}}} {M_{  \sigma\pm} ^{vc} ({\bf q})}   
{{\varphi_{\boldsymbol q}}} |^{2}}
{
{{E_{\text{exc}}}}
-\hbar \omega -i\gamma} \right]
\end{equation}
Note that we have projected the optical matrix element to the polarization direction of right- ($\sigma_-$) and left-handed ($\sigma_+$) circularly polarized light, i.e. $M_{  \sigma\pm} ^{vc} ({\bf q}) =
M_x^{vc}({\bf q}) \pm i M_y^{vc}({\bf q})
$.
The nominator determines the oscillator strength and crucially depends on the TMD properties and the lattice symmetry.   The denominator defines the energetic position $E_{\text{exc}}$  of the resonances in optical spectra
 ${{E_{\text{exc}}}=E_{\text{gap}}-E_{\text{exc}}^{\text{b}}}$ 
with the electronic band gap $E_{\text{gap}}$ and the excitonic binding energy $E_{\text{exc}}^{b}$.
Furthermore, we have introduced a dephasing $\gamma$, which accounts for radiative decay of the excitonic polarization. Non-radiative channels have not been considered. They play a crucial role at higher temperatures \cite{malte}.

Evaluating Eq. \eqref{eq_wannier}, we have access to excitonic absorption spectra of random TMDs.
Figure \ref{excAbsorption} (a) shows the spectrum for the exemplary  WSe$_2$ directly comparing the strained (orange) and the unstrained situation (black). Our approach enables to extract the contribution of excitonic effects to the strain-induced shift of resonances.  Interestingly, we find that the change in the excitonic binding energy leads to a blue-shift reducing the general strain-induced red-shift, cf. the inset of  Fig. \ref{excAbsorption} (a). This can be ascribed to the smaller effective masses (Fig. \ref{excAbsorption} (b)) entering the Wannier equation through the reduced mass $\mu$ and resulting in smaller excitonic binding energies.

Now, we discuss in detail the impact of strain on all quantities appearing in the Elliot formula, i.e. the excitonic binding energies $E_{\text{exc}}^{b}$ and excitonic wave functions $\varphi_{\bf q}$ as well as the optical matrix element $M_{\sigma\pm}^{vc}({\bf q})$ and the radiative dephasing $\gamma$. The gained insights will allow us to understand the strain-induced change in the optical fingerprint of TMD materials.

\subsection*{Strain-induced change of excitonic binding energies and wave functions}

\begin{figure}[t]
  \begin{center}
\includegraphics[width=\linewidth]{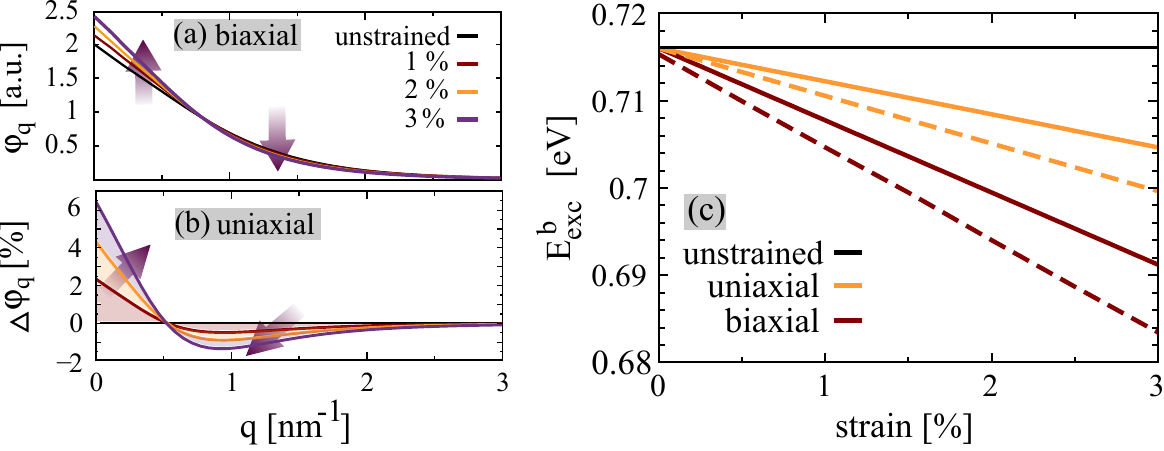} 
\end{center}
    \caption{ Strain induced changes in the (a) excitonic wave function $\phi_q$ for biaxial strain, (b)  $\Delta \varphi_{\bf q} =(\varphi_{q_x}-\varphi_{q_y})/\varphi_{q_y}$ expressing the anisotropy of the wave function for uniaxial strain, and 
(c) the excitonic binding energy for both bi- and uniaxial strain according to the solutions of the Wannier equation (Eq. (\ref{eq_wannier})) for  WSe$_2$ (solid lines) and MoS$_2$ (dashed lines).  
 The excitonic binding energy decreases both for uni- and biaxial strain due to the smaller reduced mass $\mu(s)$. We find a decrease of \unit[8]{meV} for 1\% (\unit[25]{meV} for 3\%)  applied biaxial strain for WSe$_2$ and \unit[11]{meV} (\unit[33]{meV}) for MoS$_2$.
}
  \label{excEWandEF}
\end{figure}

Strain enters in the Wannier equation (Eq. (\ref{eq_wannier})) both through the geometric and the orbital overlap effect.  From Eq. \eqref{eq_t} follows directly for the reduced mass in case of biaxial strain 
\begin{equation}
\label{eq_mu}
\mu (s) =\mu_0s^{-2}, 
\end{equation} 
where $\mu_0$ is the value for the unstrained case. For uniaxial strain the geometric effect induces an anisotropy, i.e. 
$E_{\bf q}=\frac{\hbar^2 q^2}{2\mu} \rightarrow
\frac{\hbar^2 q_x^2}{2\mu_x}
+
\frac{\hbar^2 q_y^2}{2\mu_y}
$.
Exploiting the relation $\mu_i=\mu_0s_i^{-2} $ and projecting it to elliptic coordinates with the absolute value $q$ and the angle $\phi_{ q}$, the dispersion $E_{\bf q}$ in the Wannier equation reads:
\begin{equation}
E_{\bf q}(s_x, s_y)=\frac{\hbar^2 q^2}{2\mu_{0}} \left(s_x^2 \cos^2\phi_q  + s_y^2 \sin^2\phi_q \right)
\end{equation}
Here, we have to distinguish between uni- and biaxial strain, since the solution of the Wannier equation will be different in an anistropic system. For biaxial strain with $s_x=s_y$, the Wannier equation remains isotropic and only the reduced mass $\mu$ is smaller  for tensile strain. As a result,  biaxial strain accounts for lighter and weaker bound excitons with radially symmetric excitonic wave functions. The latter become slightly larger in amplitude and spectrally narrower (Fig. \ref{excEWandEF}(a)).  
 In contrast, in  the case of uniaxial strain  the Wannier equation becomes anisotropic owing to the geometric effect. This results in anisotropic wave functions.  To quantify the degree of anisotropy, we calculate  $\Delta \varphi_{\bf q} =(\varphi_{q_x}-\varphi_{q_y})/\varphi_{q_y}$ as a function of momentum (Fig. \ref{excEWandEF}(b)). Note that we divide here by $\varphi_q^y$ (unstrained direction) to give percentage values of the strain-induced change. The anisotropy is zero for the unstrained case (black line) and increases with the applied uniaxial strain.

Beside the change in the excitonic eigen function, strain also induces a reduction of the excitonic binding energy $E_{\text{exc}}^b(s)$ due to the smaller reduced mass that can be mainly ascribed to the orbital overlap effect, cf. Eq. (\ref{eq_t}).  Our calculations reveal a decrease of $E_{\text{exc}}^b$ by \unit[4]{meV}/$\%$ applied uniaxial strain in WSe$_2$. The effect is approximately twice as large in the case of biaxial strain, cf. Fig. \ref{excEWandEF}(c).
The strain-dependent change of the excitonic binding energy scales with $s^{-2}$ according to Eq. (\ref{eq_mu}), however, at the considered low strain values $s=1.01-1.03$ the scaling is approximately linear.
The weaker bound excitons result in a blue-shift of excitonic resonances, however, this strain-induced excitonic shift is much smaller than the general red-shift of the bandgap resulting in an overall red-shift of optical resonances, cf. Fig. \ref{excAbsorption}(a). 
 The dashed lines in Fig. \ref{excEWandEF}(c) show $E_{\text{exc}}^b(s)$ in MoS$_2$, where we find qualitatively the same behaviour, however the slope of the approximately linear decrease 
is with 5 meV/$\%$ uniaxial strain quantitatively larger. The difference to WSe$_2$ can be traced back to different atomic orbital functions, where the atomic mass of molybdenum is lighter than tungsten and therefore molybendum-based TMDs tend to be generally more affected by strain.

Beside the change in the reduced mass $\mu (s)$, the strain also affects
the Coulomb matrix element $V_{\text{exc}}$ appearing in Eq. \eqref{eq_wannier}. The latter is treated as a Keldysh potential 
including a consistent description
of substrate-induced screening in quasi two-dimensional
nanostructures \cite{Keldysh1978, gunnar_prb}.  Hence, strain only enters via the tight-binding coefficients. Our calculations reveal that this effect is negligible compared to the strain-induced change of the reduced mass.
 
Our work goes beyond the discussion of strain-induced shifts, but rather focuses on the impact of strain on optical properties of TMDs that will be discussed in the following sections.

\subsection*{Strain-induced change of the optical matrix element}

The optical matrix element $\mathbf{M}^{v c}_{\mathbf k}=\langle\Psi^{v}_{\mathbf{k}}(\mathbf{r})|\mathbf \nabla|\Psi^{c}_\mathbf{k}(\mathbf{r})\rangle$ is given as the expectation value of the momentum operator $\mathbf p =-i\hbar \mathbf{\nabla}$ \cite{malic06b,carbonbuch}.
Exploiting the nearest-neighbor TB wave functions, we obtain an analytic expression for $\mathbf{M}^{v c}({\mathbf k})$
\begin{eqnarray}\label{eq_M}
{\bf{M}}^{v c}_{\bf{k}}\hspace{-3pt}=\hspace{-2pt}
c_0 \hspace{-2pt}\sum_{\alpha=1}^{3} {\bf{b}}_{\alpha} \hspace{-1pt}
(C^{v  *}_{M\bf{k}} C^{c}_{X\bf{k}} 
e^{i\bf{k} \cdot \bf{b}_{\alpha}} \hspace{-4pt} - \hspace{-2pt}
C^{v *}_{X\bf{k}} C^{c}_{M\bf{k}} 
 e^{-i\bf{k} \cdot \bf{b}_{\alpha}}\hspace{-2pt})
\end{eqnarray}
where the constant $c_0$ denotes the nearest-neighbour orbital overlap 
$
c_0=\frac{e\sqrt{3}}{a_0}\langle
\phi_{j}^{v }({\bf{r}}-{\bf{R}}_{j})
|p_x |
\phi_{i}^{c}(\bf{r}-\bf{R}_{i})
\rangle$ \cite{gunnar_prb}.
The optical matrix element exhibits, similar to the electronic band structure, a strong trigonal warping effect, i.e. it shows a triangular shape around the K and K' valleys.  This reflects the threefold symmetry of the nearest neighbors in
the real space lattice.
Note that the optical matrix element is also strongly valley dependent, i.e. at the K (K') point it is maximal, whereas it vanishes at the K' (K) point for excitation with right (left)-handed circularly polarized light \cite{gunnar_prb}. This is the microscopic origin of the observed valley polarization in TMDs \cite{cao2012valley,zeng2012valley, mak2012control}. 

\begin{figure}[t]
  \begin{center}
\includegraphics[width=\linewidth]{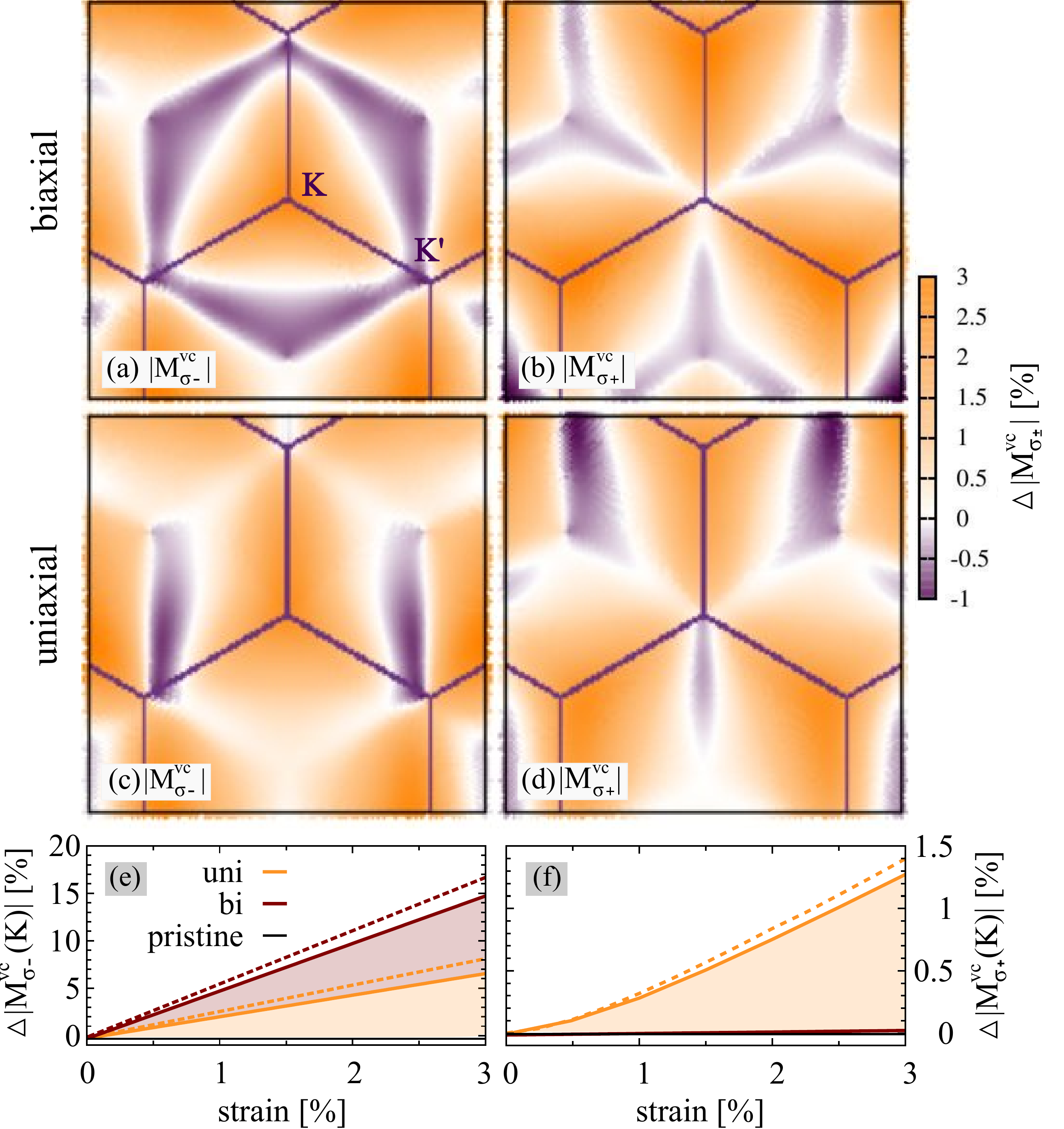} 
\end{center}
    \caption{ Influence of strain on the optical matrix element projected in the direction of right ($\sigma_-$) and left-handed ($\sigma_+$) circularly polarized light. The change of $|M^{vc}_{\sigma-}|$ and $|M^{vc}_{\sigma+}|$ are shown for biaxial [(a) and (b), respectively] and uniaxial strain [(c) and (d), respectively, for 1\% strain. Close to the K point, strain induces an increase (orange area) of the optical matrix element for both bi- and uniaxial strain. 
   Due to the valley selective excitation in TMDs, $M^{vc}_{\sigma+}$  vanishes at the K point in the unstrained case. In presence of uniaxial strain, this valley-selective optical selection rule is softened resulting in $|M_{\sigma+}| \neq 0$.  
   The change of $|M^{vc}_{\sigma-}|$ and $|M^{vc}_{\sigma+}|$ at the K point is plotted as a function of strain in  (e) and (f), respectively. 
    }
  \label{optMAtr}
\end{figure}

Strain enters in Eq. (\ref{eq_M}) through the TB coefficients $C^{\lambda}_{M(X)\bf{k}}$ and through the vectors $\bf b_\alpha$ connecting the nearest neighbors in the real space lattice. 
Figure \ref{optMAtr} (a)-(d) shows the strain-induced changes  in the optical matrix element $\Delta |M^{vc}_{\sigma\pm}| $ projected to the direction of left- or right-circularly polarized light in the case of 1 $\%$ tensile  strain. We find that strain 
has a complex impact on the optical matrix element including areas with positive (orange) and negative (purple) changes. In particular, we observe an increase of the matrix element around the K point both in uni- and biaxial case. 
This can be traced back to a large extent to the orbital overlap effect: Inserting our ansatz for the atomic orbitals into the constant $c_0$ appearing in Eq. \eqref{eq_M} we find an analytic expression for the biaxial strain:
\begin{equation}\label{eq_M2}
c_0 \propto s  \frac{2\hbar}{m_0 (\sigma_{i}^{\lambda} + \sigma_j^{\lambda})^2}.
\end{equation}
The larger the strain, the smaller are the orbital overlaps $\sigma^\lambda_i, \sigma^\lambda_j$ resulting in an increase of the optical matrix element.  

Exploiting Eq. \eqref{eq_C}
and applying a Taylor expansion of  Eq. \eqref{eq_M} around the K point, we can further evaluate the optical matrix element yielding:
\begin{eqnarray} \label{opMa}
 {\bf{M}}^{vc}_{\bf{k}}&= &
c_0 C_X^{v*} C_X^{c} i \sum_{l,m}{\bf{k}} \cdot ({\bf{b}}_{m} - {\bf{b}}_{l}) \frac{{\bf{b}}_{m}}{|{\bf{b}}_{m}|}
\\\notag
&&
\times\left(
\beta^{v*}_{\bf k}  e^{-i\fett{K} \cdot \fett{b}_{l}}e^{i\fett{K} \cdot \fett{b}_{m}} 
+\beta^{c}_{\bf k}  e^{i\fett{K} \cdot \fett{b}_{l}}e^{-i\fett{K} \cdot \fett{b}_{m}} 
\right),
\end{eqnarray} 
where we have introduced the abbreviation 
$\beta^{\lambda}_{\bf k}
= t^{\lambda}  \left(\frac{E_{\text{gap}}}{2} - E^{\lambda}_{\bf k}\right)^{-1}
$.
Neglecting the influence of strain on the tight-binding coefficients and focusing on the change of the connecting vectors (geometric effect), which are responsible for the trigonal warping effect, i.e.  ${\bf b_i} \rightarrow \mat{S} \bf b_i$, we find for the x (y) component of the optical matrix element:
\begin{eqnarray}\label{eq_M3}
 M_{x(y)}  \hspace{-4pt}\propto\hspace{-2pt} i
\frac{
s_xk_x(b_m^x-b_l^x) +s_yk_y(b_m^y-b_l^y)
}
{\sqrt{(s_xb_m^x)^2 + (s_yb_m^y)^2}}
s_{x(y)}\, b_m^{x(y)}\hspace{-4pt}.
\end{eqnarray}
For biaxial strain we find a simply relation 
$
M_{x(y)}  = s M^0_{x(y)} 
$
where $M^0_{x(y)}$ is the unstrained optical matrix element. Here, strain has the same impact on x and y direction and so the trigonal warping effect is fully conserved, cf. Fig. \ref{optMAtr}(a)-(b). 
For uniaxial strain, Eq. \eqref{eq_M3} shows that applying the strain in one direction affects the x and y components of the optical matrix element in different ways, which results in a distorted trigonal warping effect, cf. Fig. \ref{optMAtr}(c)-(d).
This feature can  be clearly traced back to the geometric effect of the strain.

Now, we evaluate the  strain-induced changes of both $|M^{vc}_{\sigma -}|$ and $|M^{vc}_{\sigma +}|$ directly at the K point, cf Fig. \ref{optMAtr}(e)-(f).  We find a linear increase of $|M^{vc}_{\sigma -}|$ with strain, where the slope is larger in the case of biaxial strain: by applying 1\% strain our calculations reveal an increase of $|M^{vc}_{\sigma -}|$ by 2$\%$ (2.5\%) for uniaxial and 4$\%$  (5\%) for biaxial in WSe$_2$ (MoS$_2$). We find again that the effect is slightly more enhanced for MoS$_2$.
Due to the optical valley-dependent selection rules,  $|M^{vc}_{\sigma +}|$ is zero at the K point in unstrained TMDs. However, in the case of uniaxial strain, we observe a softening of this selection rule due to the broken symmetry stemming from Eq. \eqref{eq_M3}. Even though the effect is rather small (increase by 0.2- 0.3 $\%$ per 1 $\%$ applied strain in WSe$_2$ and MoS$_2$), it shows that strain can be principally exploited to control the valley polarization.

To further discuss this behavior, we find an analytic expression for the optical matrix element directly at the K and K' point by evaluating the sums in Eq. \eqref{eq_M3} and directly plugging in the coordinates of the these points:
\begin{eqnarray}
\begin{pmatrix} M_x \\ M_y \end{pmatrix} \propto
\begin{pmatrix} s_x \\ \pm i s_y \end{pmatrix},
\end{eqnarray}
where $+(-)$ denotes the K (K') point.  
For the projected optical matrix element $M^{vc}_{\sigma\pm } = M_x \pm iM_y$ follows:
\begin{eqnarray}
&\quad \text{biaxial} &\quad \text{uniaxial} \notag \\
M^{vc}_{\sigma + }(K) \propto s_x-s_y  \quad & 0 &\quad \neq 0 \label{MamK}\\
M^{vc}_{\sigma - }(K) \propto  s_x+s_y \quad &\text{increase} &\quad \text{increase} \notag\\
M^{vc}_{\sigma + }(K') \propto  s_x+s_y  \quad&\text{increase} &\quad\text{increase} \notag\\
M^{vc}_{\sigma - }(K') \propto  s_x-s_y \quad & 0 &\quad \neq 0 \notag
\end{eqnarray}
Here, we clearly see that the optical valley-dependent selection rule does not apply anymore in the case of uniaxial strain.  Figure \ref{optMAtr}(f)  shows the strain-induced change of $|M^{vc}_{\sigma +}|$ at the K point. For unstrained TMDs and for biaxial strain, $|M^{vc}_{\sigma +}|$ is zero according to the optical selection rules. Applying uniaxial strain,  
$|M^{vc}_{\sigma +}|$ increases with strain as predicted in Eq. \eqref{MamK}. 
However, the effect is rather small and although it softens the optical selection rules, it will be difficult to observe the decreased valley polarization at experimentally accessible strains. The effect also occurs in MoS$_2$ (dashed line), where it is slightly more enhanced due to the larger overlaps of atomic orbital functions. 

In summary, the optical matrix element increases around the K point as a function of applied strain. This can be traced back to  the orbital effect which leads to an increase of $c_0$ (Eq. \eqref{eq_M2}) and the geometric effect changing the nearest-neighbour vectors ${\bf b}_\alpha$ and accounting for the distorted trigonal warping effect  and the softening of valley-dependent optical selection rules in the case of uniaxial strain.
As the optical matrix element directly enters the Elliot formula (Eq. \eqref{eq_elliot}), the observed changes will have a direct impact on the the optical absorption spectra. However, to fully understand the change in the oscillator strength of excitonic resonances, we  investigate the influence of strain on the excitonic linewidth.

\subsection*{Strain-induced change of the radiative dephasing}
The excitonic linewidth is expressed by the dephasing constant $\gamma$ in Eq. (\ref{eq_elliot}). 
We focus here on the  radiative decay, which is known to be the dominant dephasing channels at low temperatures, while at higher temperatures,  phonon-induced non-radiative decay channels become important \cite{malte}. The impact of strain on these processes is beyond the scope of this work. 

\begin{figure}[t]
  \begin{center}
\includegraphics[width=\linewidth]{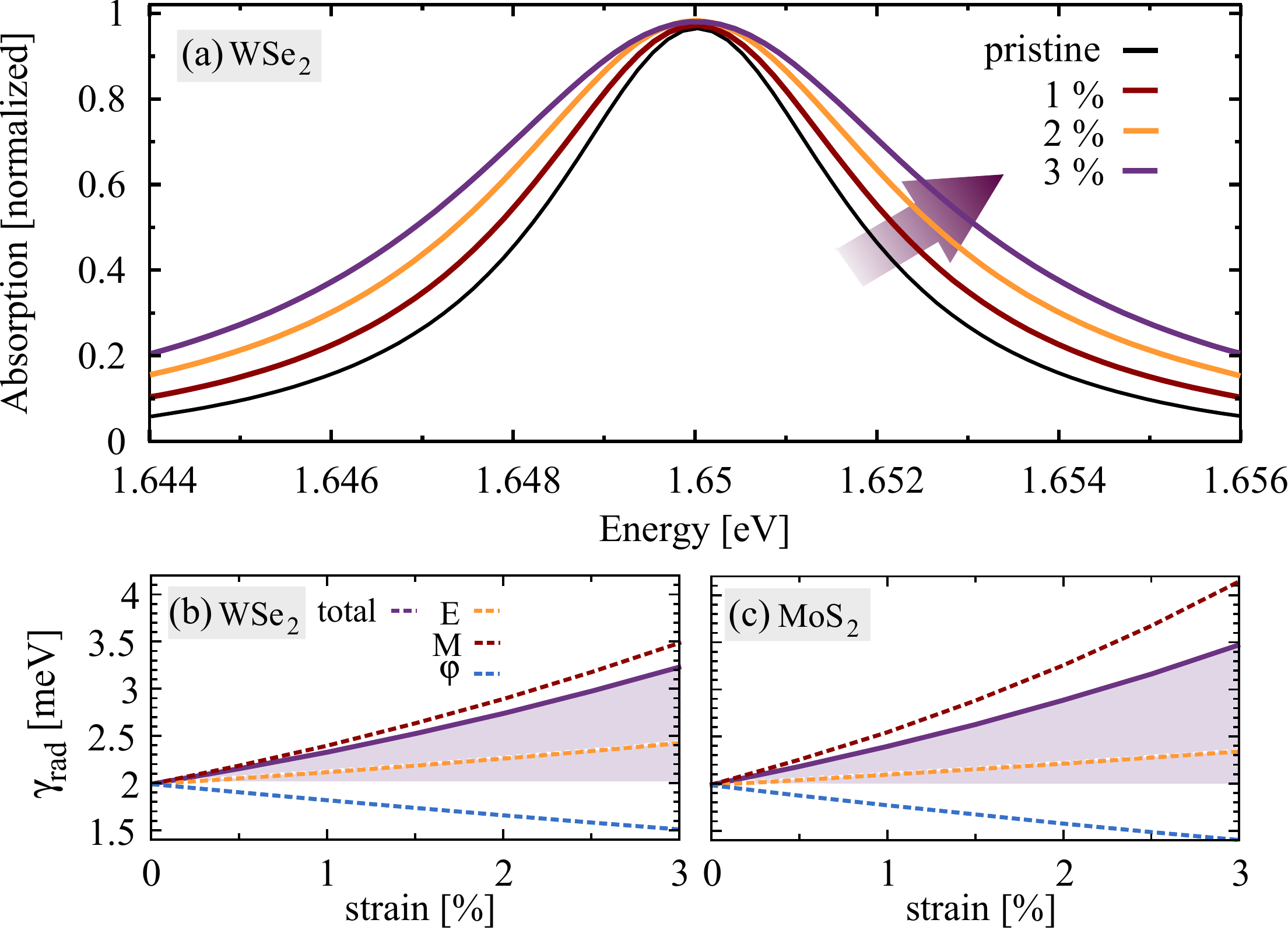} 
\end{center}
    \caption{Strain-induced broadening of the radiative linewidth. (a) Absorption spectra of strained WSe$_2$ for 1-3 $\%$ of biaxial strain.  To focus on the linewidths, the excitonic resonances are normalized and shifted to the peak in the unstrained case. We observe a clear strain-induced increase of the radiative linewidth.
Quantitative evaluation of the strain-dependent change in (b) WSe$_2$ and (c) MoS$_2$. 
We find an increase from \unit[2]{meV} in the unstrained case to \unit[2.3]{meV} (\unit[2.4]{meV}) for 1\% biaxial strain in WSe$_2$ (MoS$_2$). 
The total broadening (solid purple line) is due to the strain-induced change of the excitonic energy (dashed orange), the optical matrix element (dashed red), and the excitonic wave function (dashed blue), cf. Eq. (\ref{eq_radiative}).
The matrix element turns out to have to play the predominant role for the observed broadening.
}
  \label{gamma}
\end{figure}

The radiative decay rate is determined by spontaneous emission of light through recombination of carriers and has been  obtained by self-consistently
solving the Bloch equation for the excitonic polarization and
the Maxwell equations in a 2D geometry \cite{malte}
\begin{equation}
\gamma = \frac{\hbar^3 c \mu_0}{E_{\text{exc}} n} 
\left|
\sum_{\bf q} M^{vc}_{\sigma\pm}({\bf q}) \varphi_{\bf q}
\right |^2
\label{eq_radiative}
\end{equation}
with $\frac{c}{n}$ describing the light velocity in the substrate material and $\mu_0$ the vacuum permeability. As discussed in the previous section, the optical matrix element increases with strain suggesting an enhanced $\gamma$ as a function of strain. However, the appearing excitonic wave function and the excitonic resonance also have an influence on the final broadening. Figure \ref{gamma}(a) demonstrates that the radiative linewidth generally increases with biaxial strain in WSe$_2$.
We find an enhancement from from \unit[2]{meV} in the unstrained WSe$_2$ (black line) to \unit[3.2]{meV} for 3$\%$ biaxial strain (purple line), cf. Fig. \ref{gamma}(b). We obtain a very similar behavior for MoSe$_2$ (Fig. \ref{gamma}(c)) and uniaxial strain (not shown), where the increase of the broadening is smaller (\unit[2.6]{meV} and \unit[2.7]{meV} for 3$\%$ uniaxial strain in WSe$_2$ and MoS$_2$, respectively).  

 To get a deeper understanding of the underlying microscopic processes, we evaluate the increase of the radiative linewidth by considering separately the strain-induced changes in (i) the excitonic energy $E_{\text{exc}}$, (ii) the optical matrix element $M^{vc}_{\sigma\pm}(\bf q)$, and (iii) the excitonic wave function $\phi_{\bf q}$, cf. the dashed lines in Figs. \ref{gamma}(b)-(c) for WSe$_2$ and MoS$_2$, respectively.
Our calculations demonstrate that the optical matrix element plays the crucial role for the observed general increase of the radiative linewidth. In contrast, the excitonic wave functions actually reduce the radiative decay due to the strain-induced spectral narrowing of the wave functions (Fig. \ref{excEWandEF}(a)). Finally, the excitonic energy becomes smaller in presence of strain resulting in a larger radiative broadening (Eq. (\ref{eq_radiative})), however the effect is relatively small compared to the impact of the optical matrix element.

In conclusion, we have presented microscopic insights into the impact of uni- and biaxial strain on the optical fingerprint of atomically thin transition metal dichalcogenides. Combining Wannier and Bloch equations with the nearest-neighbor tight-binding approximation, we derive analytic expressions for the strain-induced change in the (i)  effective masses giving rise to a reduction in the excitonic binding energy, (ii) optical matrix element resulting in a softening of valley-dependent optical selection rules in the case of uniaxial strain, and (iii) radiative broadening of the excitonic resonances. We trace back these features to changes in the lattice structure (geometric effect) and in the orbital functions (overlap effect). 
The presented framework could also be extended to few-layer or encapsulated TMDs. The gained insights contribute to a better understanding of how strain changes the excitonic properties determining the optical fingerprint of these technologically promising nanomaterials. \\

This project has received funding from the European Union’s Horizon 2020 research and innovation programme
under grant agreement No 696656 (Graphene Flagship), the Chalmers Area of Advance Nanoscience  and Nanotechnology, and the Swedish Research Council (VR). Furthermore, we are grateful to Andreas Knorr (TU Berlin), Rudolf Bratschitsch, Steffen Michaelis de Vasconcellos, Robert Schmidt, and Iris Niehues (University of Muenster) for inspiring discussions on the impact of strain on 2D materials.



%

\end{document}